# Non-Hermitian engineering of single mode two dimensional laser arrays

*Mohammad H. Teimourpour[1,2], Li Ge[3,4], Demetrios N. Christodoulides[5] and Ramy El-Ganainy[1,2,*]*

[1]Department of Physics, Michigan Technological University, Houghton, Michigan, 49931, USA

[2]Henes Center for Quantum Phenomena, Michigan Technological University, Houghton, Michigan, 49931, USA

[3]Department of Engineering Science and Physics, College of Staten Island, CUNY, Staten Island, NY 10314, USA

[4]The Graduate Center, CUNY, New York, NY 10016, USA

[5]College of Optics & Photonics–CREOL, University of Central Florida, Orlando, Florida 32816, USA

[*]ganainy@mtu.edu

A new scheme for building two dimensional laser arrays that operate in the single supermode regime is proposed. This is done by introducing an optical coupling between the laser array and a lossy pseudo-isospectral chains of photonic resonators. The spectrum of this discrete reservoir is tailored to suppress all the supermodes of the main array except the fundamental one. This spectral engineering is facilitated by employing the Householder transformation in conjunction with discrete supersymmetry. The proposed scheme is general and can in principle be used in different platforms such as VCSEL arrays and photonic crystal laser arrays.

**Introduction**

Over the last two decades, vertical cavity surface emitting lasers (VCSEL's) have become a mature technology that finds several different applications [1,2]. The output power of a VCSEL laser is proportional to the area of its active region. Increasing the emission power can be thus achieved by building large area VCSELs. This however introduces some difficulties such as filamentation [3] and multimode operation [4-7]. An attractive alternative is to build an array of coupled VCSEL lasers. If all the elements in the array lase in the same phase, the output intensity at the focus of the emitted laser beam can be enhanced by a factor proportional to the square of the number of the individual elements constituting the array. Today, VCSEL laser arrays can provide up to several watts of output powers and play an important role in many applications in industry, communication networks, and multimedia to just mention a few [1]. Despite this tremendous success, laser arrays in general suffer from multimode operation. While several techniques for eliminating longitudinal modes of individual resonators exist [8-11], suppressing the transverse supermodes is by no means an easy task [3-6]. This can be explained by noting that these supermodes arise as a result of eigenfrequency splitting introduced by the coupling between the individual laser resonators. Given that the discrete spectral band associated with these supermodes is usually smaller than (or comparable to) the spectral gain bandwidth of typical semiconductor active media, lasing competition between these eigenmodes takes place. This can lead to a multimode operation that affects the stability of the laser output power as well the quality of the emitted laser beam [12]. While nonlinear gain saturation might favor some modes over others, a lack of precise control over the operation of these laser systems can present a practical challenge.

Recently, a technique based on discrete supersymmetry (DSUSY) [13-17] was proposed in order to solve similar problems for one dimensional (1D) laser waveguide arrays [18]. In that work, a

pseudo–isospectral array, generated through DSUSY, was employed as a reservoir that spoils the quality factors of all higher order transverse supermodes while at the same time leaving the fundamental supermode intact. This was possible primarily owing to a particular feature of DSUSY: the effective Hamiltonian matrix describing a 1D array with nearest neighbor coupling (NNC) is symmetric and tridiagonal and can be thus used with DSUSY to engineer a pseudo-isospectral reservoir [18]. Despite the appeal of this technique, it cannot be applied for two dimensional (2D) laser arrays for the following reason: unlike their 1D counterparts, 2D arrays cannot not be represented by a tridiagonal matrix. Thus, blindfold application of 1D DSUSY to the effective Hamiltonian matrix of a 2D array results in a reservoir with complicated higher order couplings beyond NNC, which is not practical to implement. Thus, the following question naturally arises: can one find a variant of 1D DSUSY that can be applied to 2D systems? In other words, can we engineer an implementable pseudo-isospectral reservoir that has an identical spectrum of a 2D laser array except for the fundamental supermode, in order to eliminate the multimode dynamics and achieve only single mode laser emission (as depicted schematically in Figure.1)?

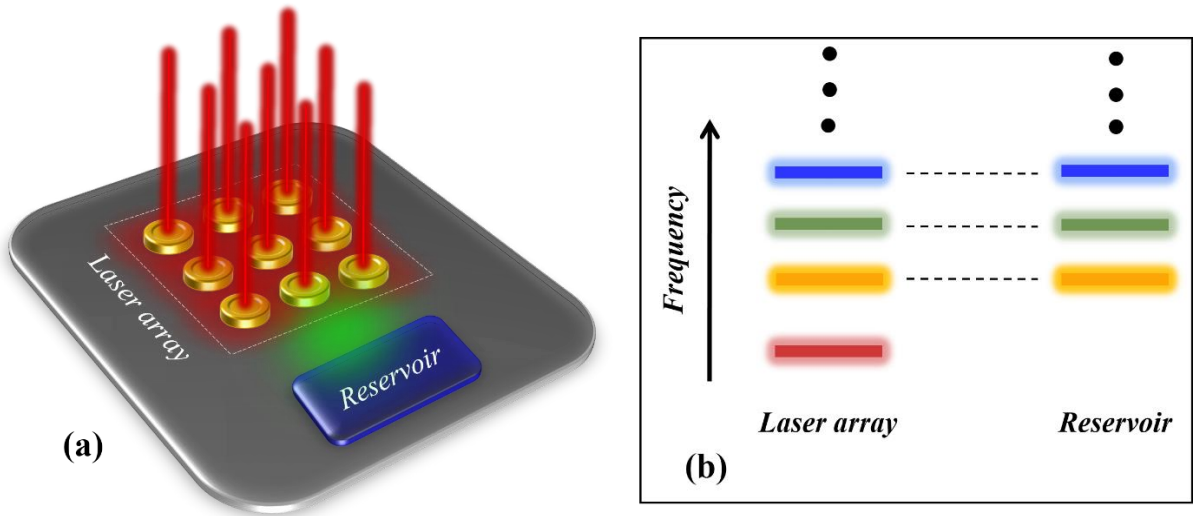

Figure 1: (a) Schematic illustration of a 2D laser array coupled to an optical reservoir that is engineered to suppress higher order modes and that allows only for single mode operation. (b) Eigenvalue spectra of the laser array and the reservoir. All the eigenfrequencies of both subsystems are matched except the fundamental one (indicated by red thick line). By introducing an optical coupling between the laser array and the reservoir, it is possible to suppress all the higher order lasing modes.

In this work we provide an answer to this question by using the Householder transformation method [19] in conjunction with discrete supersymmetry [18]. In particular, starting from the banded matrix that describes the 2D laser array, we first use the Householder method in order to obtain an isospectral tridiagonal matrix that can be implemented by a 1D chain of coupled resonators with only NNC. Note that due to the spectral degeneracies of 2D systems, and since degeneracy is not permitted in 1D systems [20] (see section A of the supplementary material), the resultant isospectral 1D array will in general consists of a union of two different isolated chains. In terms of matrices, this means that applying the Householder transformation to a banded matrix that represents a 2D array will yield a tridiagonal matrix that exhibits several block diagonals, each of which has a non-degenerate spectrum. The next step is to apply DSUSY transformation to the 1D chain in order to create an auxiliary structure that shares all the eigenfrequencies of the original laser array except that associated with the fundamental supermode. By introducing optical losses to this 1D discrete reservoir (consisting of several independent 1D chains) and by engineering its coupling to the original 2D laser array, the quality factors of the higher order supermodes of the main 2D laser array are spoiled and as a result, their lasing thresholds are increased, all while leaving the lasing properties of the fundamental mode (threshold and profile) almost intact. Finally, by using the results of the last step as an initial starting point and by considering laser rate equations and assuming fast carrier dynamics, we optimize the design in order to achieve the desired performance under gain saturation nonlinearity−a model that has been shown to yield good results for analyzing PT symmetric lasers [21,22]. These steps are summarized by the flowchart shown in Figure. 2.

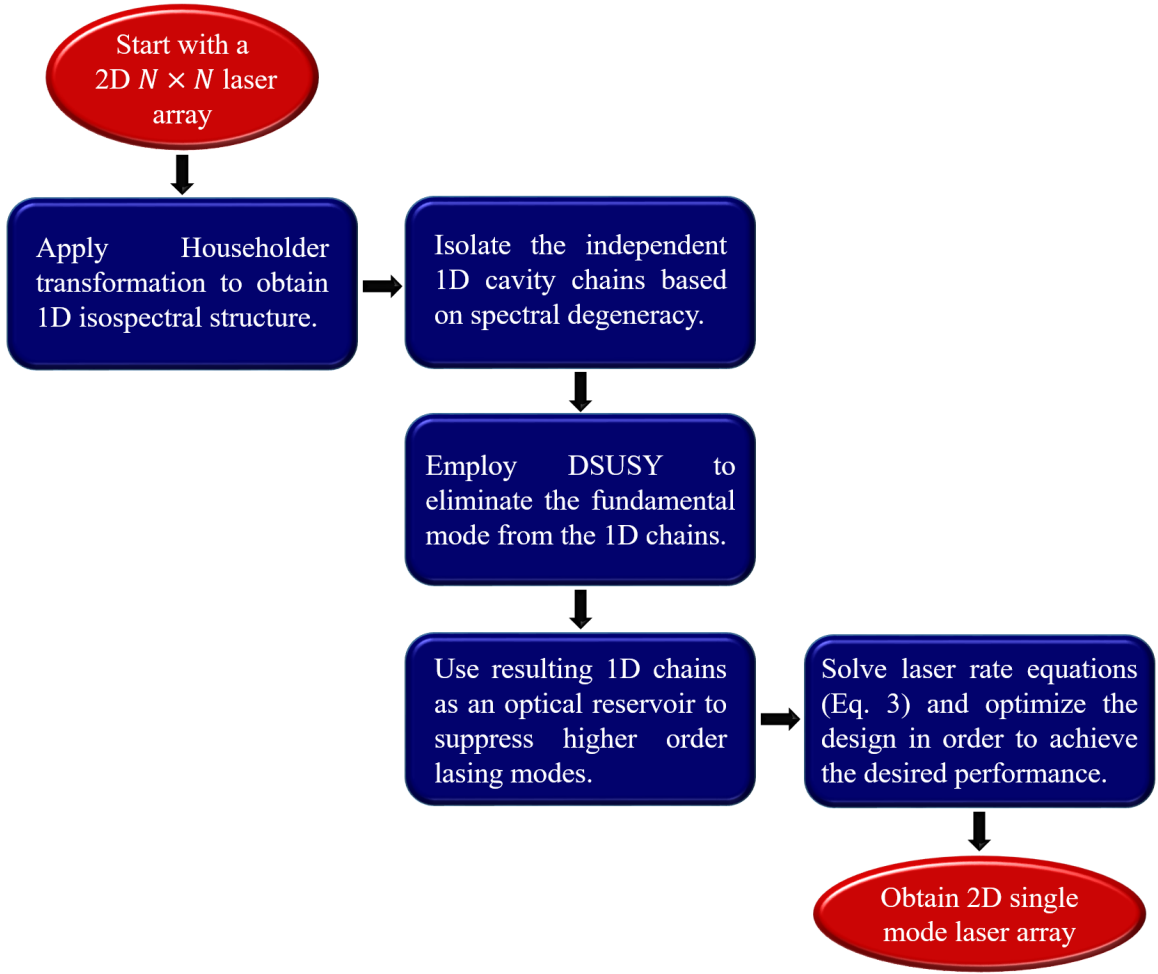

Figure 2: A flowchart that illustrates the main steps for designing 2D single mode laser arrays based on Householder transformation in conjunction with discrete supersymmetry techniques.

**Single mode operation in uniform square laser arrays**

The Householder-DSUSY technique described in the introduction is very general and can be applied to any array topology. However, here we focus on square arrays due to their simple practical feasibility. In particular, we deal in this section with uniform square arrays having $N \times N$ optical cavity elements. Within the framework of coupled mode theory and before or just at lasing threshold where the system is linear:

$$i\frac{da_{m,n}}{dt} = (\omega_o - i\gamma + ig)a_{m,n} + \sum_{m',n'} \kappa_{m,n}^{m',n'} a_{m',n'} \qquad (1)$$

In equation (1), $a_{m,n}$ is the optical field amplitude in the cavity element characterized by the integer indices $m$ and $n$, $\omega_o$ is the uniform resonant frequency resonant modes of each isolated optical cavity, $\gamma$ and $g$ represents the optical loss (radiation and material ) and gain, respectively while $\kappa_{m,n}^{m',n'}$ are the elements of the coupling matrix which are finite only for next neighboring coupling and zero otherwise and $t$ represents time. In the model described by equation (1), we have neglected diagonal coupling between the resonators as well as other higher order couplings beyond NNC. This treatment is justified, for example, for closely packed cylindrical VCSEL arrays, with the radius $r$ of each cavity being much larger than their spacing $d$. As such, the diagonal distance between two cavities $2(\sqrt{2}-1)r + \sqrt{2}d$ is also much larger than $d$, indicating a negligible diagonal coupling when compared with NNC. We note however that our proposed method can be still applied without any restrictions if higher order couplings are taken into account.

Equation 1 can be recast in a Hamiltonian form:

$$i\frac{d\vec{a}}{dt} = H\vec{a}, \qquad (2)$$

where $\vec{a} = [a_1 \ a_2 \cdots a_l \cdots a_{N^2}]^T$ is the optical amplitude vector and $T$ denote matrix transpose. The subscript $l$ in $a_l$ is related to the double indices notation by $l = N \times (m-1) + n$. Note that, written in a frame rotating with an angular velocity of $\omega_o$, the diagonal elements of the Hamiltonian matrix $H$ contain only imaginary numbers that represent the optical loss of each cavity. As we have discussed earlier, since $H$ describes a 2D array, its structure is not tridiagonal.

Thus the DSUSY method used in ref. [18] cannot be applied here to achieve single mode lasing. However, this difficulty can be overcome by first applying the Householder transformation to generate a tridiagonal matrix $\hat{H}$, which is isospectral to $H$ [23] and can be then used in conjunction with DSUSY to engineer the quality factors of the higher order modes in order to ensure single mode operation. We note that the Householder method has been employed recently for studying eigenvalue statistics in networks with reduced dimensions [23].

Based on our discussion so far, one might anticipate that the reservoir should at least consist of $N^2 - 1$ optical resonators in order to "kill" all the higher order lasing eigenmodes. However, this is not necessarily correct since, unlike 1D arrays, 2D structures can exhibit spectral degeneracies. In particular, as we show in supplementary material, the spectrum of a 2D $N \times N$ uniform square array consists of $(N^2 + 2)/2$ distinct eigenvalues when $N$ is even and $(N^2 + 1)/2$ distinct eigenvalues when $N$ is odd. As we will see, this feature can be employed to simplify the reservoir structure. Finally, we note that in the absence of any applied gain, the combined laser and reservoir system is described by an equation similar to (2) except that the amplitude vector now contains the field amplitudes in the laser array as well as the reservoir chains. Additionally, the Hamiltonian of the total system $H_{SR}$ will also contain information about the system, reservoir and their interactions. If the reservoir consists of $M$ resonators, $H_{SR}$ will have $N^2 + M$ different eigenmodes, each of which will vary as $\exp(-i\Omega_p t)$ with $p = 1, 2, ...., N^2 + M$. The real part of $\Omega_p$ represents the oscillating frequency while its imaginary part indicates the modal loss coefficient. As gain is applied to the system, the imaginary parts of the eigenvalues $\Omega_p$ will be in general shifted upward and the mode with the minimum imaginary part will start to lase first. Without the reservoir, even when the fundamental mode is the first mode to lase, other modes follow suit shortly above its threshold, resulting in multi-mode lasing instead of single-mode lasing. The role of the reservoir thus is to shift down the imaginary parts of all $\Omega_p$ before the gain is applied ("Q-spoiling"), except the fundamental mode. For simplicity, we will denote the complex eigenfrequency of that fundamental mode by $\Omega_f = \Omega_R + i\Omega_I$, where $\Omega_I < 0$ before lasing and $\Omega_I = 0$ at the lasing threshold.

In what follows we illustrate the power of our proposed method for eliminating the higher order lasing modes by considering two different examples of 2D uniform square laser arrays that consist of $3\times 3$ and $4\times 4$ coupled lasing elements respectively.

It is important to note that in all our studies we assume that only the optical resonators that comprise the main laser array are pumped. This configuration can be easily achieved in systems that rely on either optical pumping [24] or electric pumping [25].

Finally, and before we move to specific numerical examples, we note that in all the following, we have rounded all the resultant numbers for the frequencies and coupling coefficients that results from the procedure described in figure 2 to the first decimal points whenever appropriate.

**Example I: 3 × 3 laser array:**

Here we consider a square laser array consisting of nine identical resonators arranged on a square grid. At the lasing threshold, the system is linear and can be described by equation (2). We focus here on uniform arrays and we assume a normalized value for the coupling and loss coefficients of $\kappa_{mn} = \kappa = 1$ and $\gamma = 0.1$, respectively. This structure exhibits five distinct eigenvalues (see supplementary material) and thus, in principle only four auxiliary resonators are required in order to achieve single mode lasing. By applying the general recipe for achieving single mode operation as discussed above (see figure 2), we obtain two independent, one dimensional chains of optical resonators (one of them consists of just one element). However, by investigating the resultant structure, we found that it is necessary to include an additional resonator in order to achieve single mode operation. This can be attributed to the fact that matching the eigenfrequencies of the laser array and the reservoir is not a sufficient condition to "kill" all the undesired states. The spatial overlap between the eigenmodes must also be taken into account.

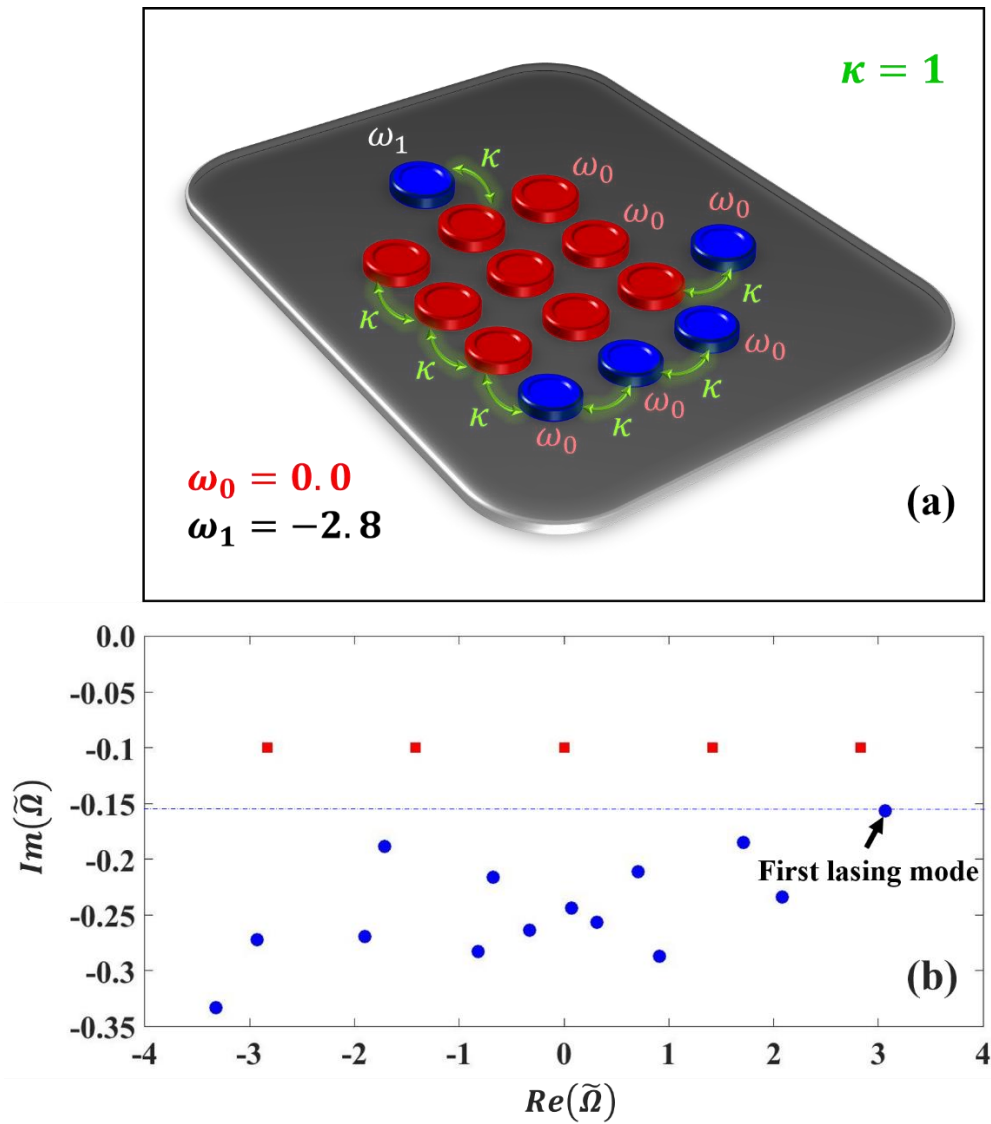

Figure.3: (a) Schematic illustration of a 3×3 laser array (red) coupled to a reservoir that consists of five optical cavities (blue). All the eigenfrequencies of the individual photonic resonators and the coupling coefficients (which are all identical) are depicted on the same figure. Note the procedure described in figure

2 yields only the reservoir chain at the bottom and the single resonator at the top. The extra cavity at the right side is added to optimize the performance. (b) Eigenvalue spectrum of the laser array alone in the absence of the reservoir (square red marks) as well as that of the combined system (blue circle marks). In (b) $\tilde{\Omega} = \Omega_p / \kappa$, and as mentioned in the text $\omega_0$ was taken to be zero. We have assumed that the optical loss coefficient of each cavity in the reservoir to be $\gamma_R = 0.5$.

By including the additional resonator (located to the right of the main array in figure 3 (a)) and carefully engineering the interaction between the reservoir and the main array, the fundamental mode indeed has the highest quality factor as evidenced by figure 3 (b), which indicates that it will lase first as the pump power increases. This is in contrast to the spectrum of the isolated array that shows that all modes will have the same lasing threshold. Thus based on our linear analysis, we indeed anticipate that our proposed structure will result in single mode operation within certain range of applied gain values. Interestingly, we also note that though the linear analysis indicates that the system will function as expected, the reservoir does not leave fundamental mode completely intact but rather shifts its complex eigenfrequency. This can be understood by noting that our recipe as described in figure 2 neglects off-resonant interactions [18] which is responsible for this shift. However, this is a smaller effect compared to the resonant coupling and thus does not affect the effectiveness of our approach. It important to note that the non-Hermitian engineering with additional loss in the reservoir only slightly perturbs the phase coherence of $a_l$ in the fundamental mode, with a maximum phase deviation $0.02\pi$.

Before we move to the next example, we note that normalized value $\omega_1 = -2.8$ in the rotating frame means real value $\omega_1 = \omega_o - 2.8\kappa$ in the nonrotating frame. Similar argument applies to all of the following discussions.

**Example II: 4 × 4 laser array:**

Having demonstrated our technique for a laser array that consists of nine elements, we now show that this method can be also applied to even larger systems by considering a $4 \times 4$ laser array. In this case, there are sixteen different eigenmodes, of which only nine are distinct (see supplementary material) and we find that a reservoir that consists of eight optical resonators is sufficient to suppress the higher order modes. By following the aforementioned procedure as described in figure 2, we obtain the structure shown in figure 4. Here, the optical loss coefficient is assumed to be $\gamma = 0.1$ for the main laser array and $\gamma_R = 0.8$ for the reservoir chains.

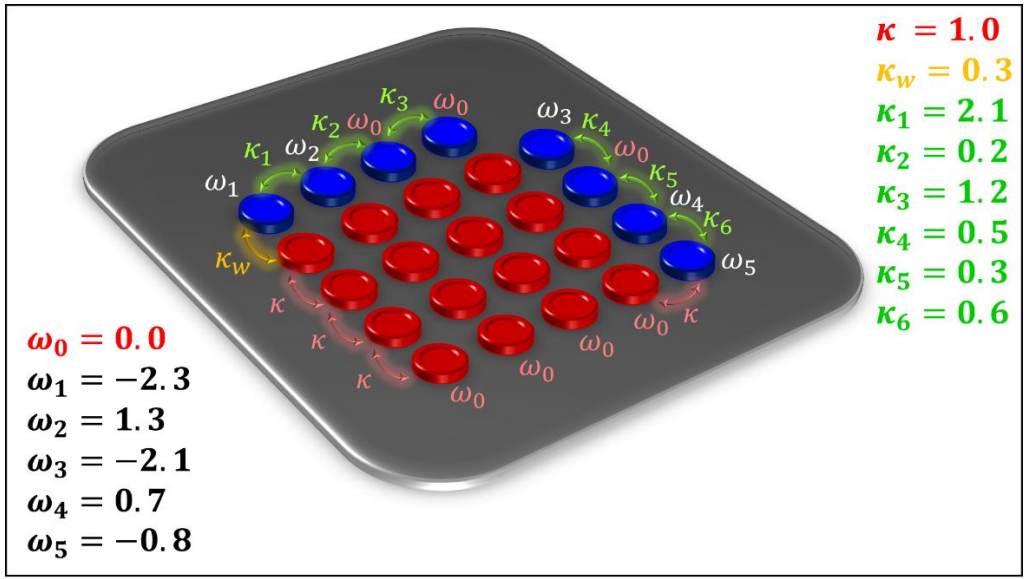

Fig.4. Schematic representation of a $4 \times 4$ laser array (red cavities) coupled to a reservoir consisting of eight resonators (blue cavities). The optical losses of the laser cavities and the reservoir are assumed to be $\gamma = 0.1$ and $\gamma_R = 0.8$, respectively. The normalized optical frequencies in the rotating frame are shown in the figure. The coupling profile of the reservoir and its interaction with the main array is also described in the same figure.

Similar to the previous scenario of nine elements array, here the eigenvalue distribution (not shown) also indicates that the fundamental mode exhibits the highest quality factor and thus will start to lase before any other mode as gain is applied.

An interesting observation here is that neither the structure of the reservoir nor its coupling profile to the main array are unique. In particular, due to the degeneracy of the main array, several different combinations for the reservoir chains can be obtained. Furthermore, for each reservoir structure, the coupling profile can be engineered in different fashions. For example in figure 4, the reservoir chains are assumed to be parallel to the main laser array with a uniform coupling coefficient $\kappa$ and $\kappa_w$. Alternatively, one could connect the chain in a number of different ways or even use similar layout but with different coupling coefficients. Obviously this is an optimization problem with the optimal design defined by maximizing the single mode operation regime. Our strategy here is to demonstrate the basic idea and we carry out the optimization investigations in a different work. Throughout this work, the optimized designs were obtained by using the initial results of Householder and SUSY transformation in conjunction with trial and error parameter scanning. Employing specialized optimization algorithms is expected to yield devices with better performance and is a topic that we explore elsewhere.

**Bosonic-inspired two dimensional laser arrays**

In the previous examples of uniform square arrays, we have seen how the spectral degeneracy led to a simplified reservoir design with less optical resonators than those would have been needed in the absence of degeneracy. In these square uniform arrays, the multiplicity of the eigenvalues is a direct outcome of the spatial symmetry. In particular, the system exhibits a point symmetry group called $D_4$ that characterize its spectral feature.

Here we explore a different type of square arrays that do not have uniform coupling across the structure: the so called $J_x$ arrays. These interesting configurations have been discovered in the context of spin networks [26] and later mapped to optical waveguides platforms [27]. While their behavior can be explained by using the angular momentum algebra, it was shown recently that their spectra and dynamics can be also understood by using bosonic algebra [28]. While the two different approaches (spin and bosonic algebras) for investigating $J_x$ arrays are formally equivalent, the second picture proved more useful in building higher order networks that inherit most of the properties associated with the 1D $J_x$ array [29].

The most pertinent feature of 1D $J_x$ arrays in our context is their equidistant ladder of eigenvalues. Due to this property, a 2D discrete structure formed by taking the tensor product of two, 1D $J_x$ array will possess additional accidental degeneracies beyond those arising from spatial symmetries for $N > 3$. As we show in the supplementary material, the number of distinct eigenvalues in an $N \times N$ bosonic array is given by $2N - 1$. Thus for example, while a uniform $4 \times 4$ structure has nine different eigenvalues, its counterpart $J_x$ array will have only 7 different eigenvalues. This in turn can lead to a further simplification of the reservoir design. We verify this observation for a $4 \times 4$ bosonic laser array as shown in figure 5. Similar to the $3 \times 3$ case, here we find that adding an extra resonator can enhance the single mode performance and we thus end up with 7 optical cavities in the reservoir instead of 6.

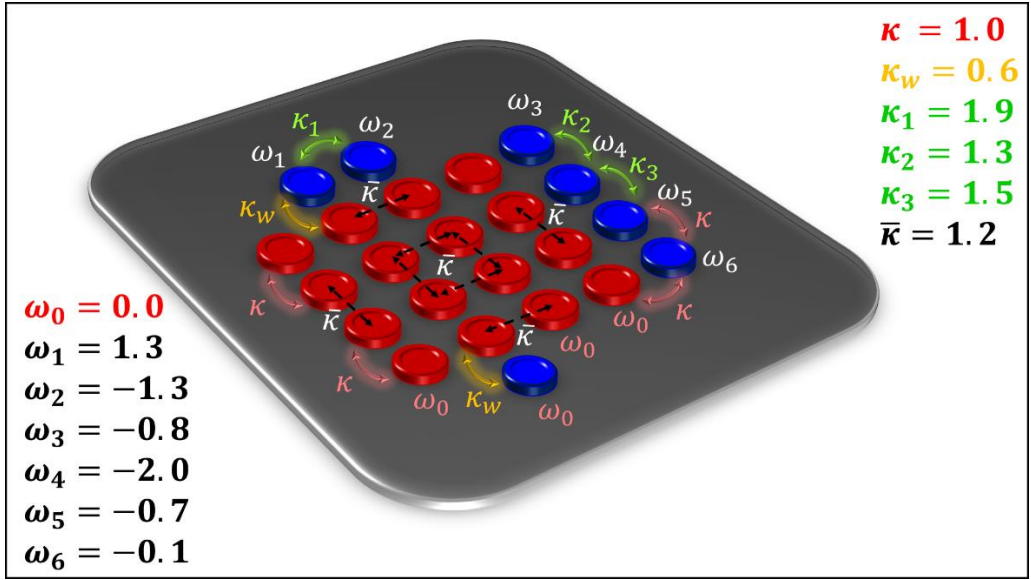

Fig.5. A schematic representation of a $4 \times 4$ bosonic $J_x$ laser array (red) interacting with a reservoir consisting of seven resonators (blue). The normalized frequencies and coupling coefficients are depicted in the figure. Note that the coupling profile of the main array is no longer uniform. Here, the optical loss coefficients of the optical resonators associated with the laser array and the reservoir are assumed to be $\gamma = 0.1$ and $\gamma_R = 0.4$, respectively.

The eigenvalue distribution of the structure shown in figure 5 reveals that in this case also only one mode will start to lase as gain is applied.

Again, similar to our previous discussion, there is more than one way of engineering the reservoir and its coupling to the main array. An optimal design thus requires carrying out optimization analysis which we consider elsewhere.

We note that building 2D bosonic arrays similar to those discussed here is not fundamentally different from building uniform arrays. In the former, the separation between the resonators has to be carefully adjusted in order to produce the desired coupling profile which has been shown to be possible in 1D [27].

Finally, it would be of interest to compare the far field emission pattern from $J_x$ and uniform laser arrays. However, this comparison cannot be carried out based on coupled mode analysis or laser rate equations, instead one has to consider a particular implementation and use full wave analysis which we plan to investigate in subsequent publication.

**Emission dynamics**

So far, we have investigate the single mode behavior of the engineered laser arrays near the lasing threshold where the system can be treated by using linear coupled mode theory. However, laser dynamics above threshold are intrinsically nonlinear [12, 30] and the lasing emission in this regime cannot be deduced from the linear analysis. Here we explore the nonlinear behavior of the laser structures proposed in this work in order to demonstrate that the single transverse mode emission persists in the existence of nonlinearity and quantify the range of single mode operation as a function of the applied gain. While different models for laser nonlinearities do exist, here we consider a scenario where the carrier dynamics are very fast and can be integrated out of the laser rate equations. We thus consider a system of equations similar to (1) except that the linear gain coefficient is replaced by a nonlinear gain saturation term:

$$i\frac{da_{m,n}}{dt} = \left(\omega_{mn} - i\gamma + \frac{i\,g}{1+\alpha|a_{m,n}|^2}\right)a_{m,n} + \sum_{m',n'} \kappa_{m,n}^{m',n'}\, a_{m',n'} \quad (3)$$

where $\alpha$ is the gain saturation coefficient. For the reservoir resonators, the coefficient $\gamma$ is replaced by $\gamma_R$ and no gain is applied. To illustrate the nonlinear emission dynamics, we consider example of $3\times 3$ laser array depicted in figure 3. By integrate equation (3) for that specific structure starting from random noise, we obtain the temporal evolution of the field intensities inside each resonator in the laser array as shown in figure 6. In this simulation, the gain coefficient was assumed to be $g = 0.4$ and the gain saturation coefficient was taken to be $\alpha = 1.0$.

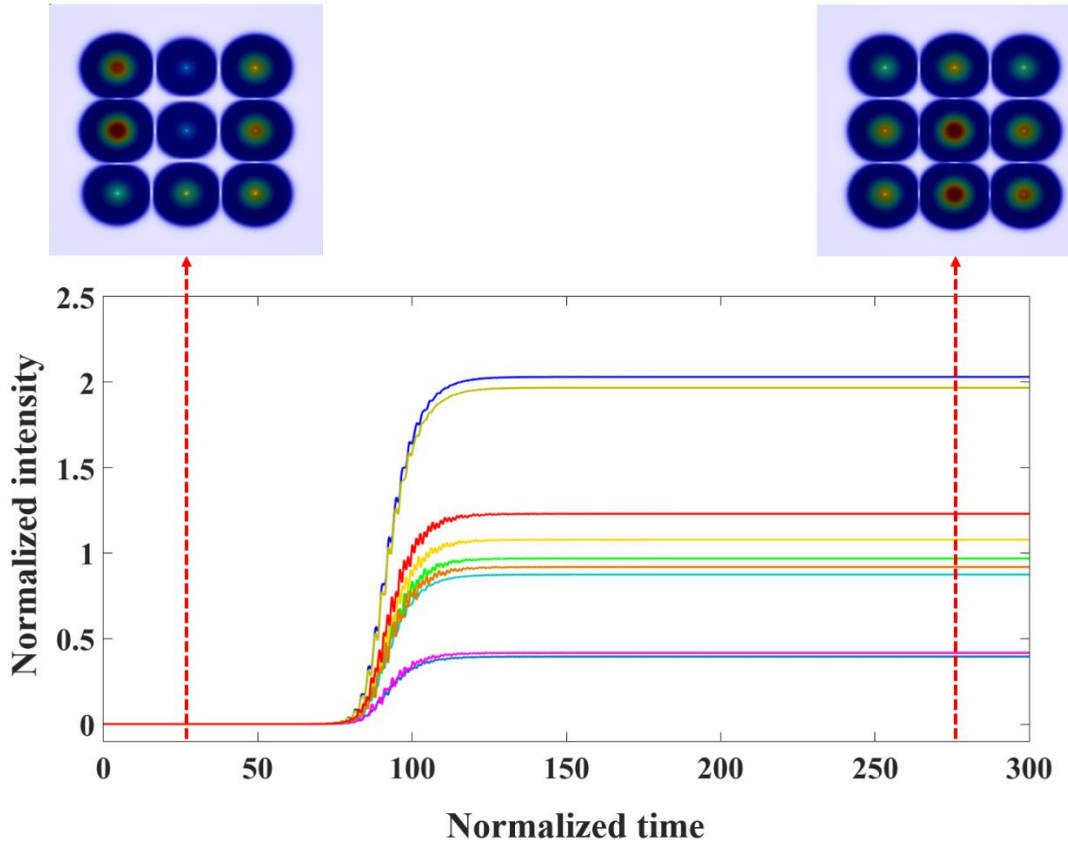

Fig.6. Nonlinear dynamics of optical intensities in the $3\times 3$ laser array shown in figure 3 with the discrete reservoir taken into account. Here we also depict the temporal snapshots of the lasing pattern $|a_{m,n}|^2$ at different times where steady state single mode emission can be observed after a transient period.

In order to confirm that lasing mode is indeed the fundamental state with all the optical fields in different resonators oscillating in phase, we have also checked the phase dynamics (not shown here) of the field amplitudes and we observed in-phase oscillations with very small differences between the phases of the individual elements, which is consistent with the eigenmode analysis we have discussed in Sec. 2.1. This feature is crucial in order to ensure that light emitted from different cavities will add constructively at the center of the far field emission plane. This in turn results in a far field bright spot that has an intensity proportional to the square of the number of elements in the array, i.e. $\sim N^4$. In practice however, due to the unequal amplitude distribution between the resonators (see figure 6), the intensity of bright spot will be less than $N^4$.

Our time-dependent simulations show that the single mode operation persists even when the gain is increased to more than two times its threshold value. To confirm these results, we use techniques similar to those employed in the Stead-state Ab-initio Laser Theory (SALT) [31,32] to calculate the steady state solutions which indeed confirms that the single mode operation extend in the range $0.18 \leq g \leq 0.5$. Interestingly, that range can be even significantly extended if the extra resonator on the right is removed and one of the coupling coefficients between the reservoir chain at the bottom is changed to $\kappa = -1$. While negative coupling is not common in microring and microdisk platforms, it can be engineered in photonic crystal cavities [33,34]. We also note that if we had chosen to pump the laser array as well as the reservoir, we find a much narrower range of single mode operating, i.e., $0.16 \leq g \leq 0.32$. This optimization process would correspond to the last step in the flowchart shown in Fig.2. We have also carried out similar simulations for the case of $4 \times 4$ arrays and Bosonic $4 \times 4$ and we have observed single mode operation for different range of the gain coefficient $0.13 \leq g \leq 0.25$ and $0.13 \leq g \leq 1.4$, respectively.

**Discussion and concluding remarks**

In this work we have proposed a new approach for building single mode two dimensional laser arrays based on engineering a discrete reservoir whose functionality is to suppress higher order lasing modes while leaving the fundamental mode almost intact. Our approach builds on our recent work for one dimensional laser arrays [18] except that here we cannot use discrete supersymmetry directly. Instead we first apply the Householder algorithm to laser array in order to generate an isospectral 1D discrete structure which can be then used in conjunction with supersymmetry to achieve single mode operation. Using the linear analysis at lasing threshold, we have shown that this technique indeed predicts single mode emission for $3\times3$ and $4\times4$ uniform arrays. We also demonstrated that using a special type of non-uniform arrays whose properties are derived by using bosonic algebra can result in a higher dimensional degenerate landscapes which in turn can lead to a simplified reservoir design. Additionally, we have confirmed the single mode operation of the proposed laser structures and studied their nonlinear dynamics under gain saturation for arbitrary initial conditions as noise. Our numerical investigations shows that a reasonable range of single mode operation is achievable in the $3\times3$ and $4\times4$ arrays.

While the approach proposed in this work can be applied in principle to any laser array system (microrings, microdisks, VCSELs or photonic crystals), general considerations must be taken into account during implementation. In particular, in order for the proposed structures to function as intended, the coupling coefficients between the different elements of the main array should be large compared to the bandwidth enhancement factor. Otherwise, the nonlinear evolution of the system can still result in chaotic emission. Another practical advantage for operating in the strong coupling regime is fabrication tolerances in the following sense. All the parameters of the auxiliary array (the reservoir) in the previous examples including the frequency detuning from $\omega_o$ (the resonant frequency of the cavities in the main array) are calculated in terms of some coupling coefficient which was taken to be unity. If this coupling coefficient is very small compared to $\omega_o$, it will require a very high fabrication precession in order to implement the new resonators with these small detuning values. On the other hand, if say $\kappa/\omega_o \in [10^{-3}, 10^{-2}]$, we can roughly estimate that for an optical cavity with a characteristic length scale (radius of microring cavity for example) of $10\,\mu m$, the necessary change in the cavity dimension to produce the required frequency detuning will be of in the range between 10-100 $nm$ which can be achieved by using today's fabrication techniques. We note that the above discussion implies a coupling factors in the order of 0.1-1 THz which has been already observed experimentally in transversely coupled VCSELs [35,36] and in photonic crystal platforms [37]. Alternatively, electric biasing can be used to fine-tune the resonators eigenfrequencies to their desired values [38]. Furthermore we note that the non-uniformity in the physical parameters (resonant frequencies, coupling coefficients, etc.) due to limited fabrication precision can be an important factor in determining the device performance. In our previous study of 1D SUSY laser arrays [18], we have shown that even with 10% disorder, single mode operation can be still achieved. We have confirmed here also that in the 2D case, single mode operation can be still obtained for different realizations with 10% disorder in design parameters. Finally we comment on the scalability of our approach. For any $N \times N$ square array (uniform or bosonic), the total number of resonators is $N^2$ but the number of resonators on the four edges is $4N-4$. In the case of uniform arrays, despite the fact that the number of supermodes having distinct eigenvalues scales as $\sim N^2/2$ for large $N$ (see section B of the supplementary material), these modes can be still accessed and influenced by the reservoir since they are extended all over the array with finite intensity contribution at the edge elements. In the case of bosonic array, the number of modes with different eigenfrequencies scales as $2N$ and can be also accessed via the edge resonators. However it is important to note that for large $N$, disorder induced localization can render some of the modes inaccessible through the edge cavities- a limitation that persists in all systems consisting of large number of coupled oscillators. It is thus clear that this current work presents several opportunities as well as challenges that merit more theoretical and experimental investigations which we plan to carry elsewhere.

**Acknowledgement**
R. E. acknowledge support under NSF Grant No. ECCS-1545804 and support from Henes Center for Quantum Phenomena at Michigan Technological University. L.G. acknowledges support under PSC-CUNY Grant No. 68698-0046 and NSF Grant No. DMR-1506987.

# Supplementary material

# Non-Hermitian engineering of single mode two dimensional laser arrays

*Mohammad H. Teimourpour[1,2], Li Ge[3,4], Demetrios N. Christodoulides[5] and Ramy El-Ganainy[1,2,\*]*

[1]Department of Physics, Michigan Technological University, Houghton, Michigan, 49931, USA

[2]Henes Center for Quantum Phenomena, Michigan Technological University, Houghton, Michigan, 49931, USA

[3]Department of Engineering Science and Physics, College of Staten Island, CUNY, Staten Island, NY 10314, USA

[4]The Graduate Center, CUNY, New York, NY 10016, USA

[5]College of Optics & Photonics–CREOL, University of Central Florida, Orlando, Florida 32816, USA

[\*]ganainy@mtu.edu

**A. Householder Method**

Householder method for tridiagonalization of matrices was introduced by A. S. Householder in 1985 [1]. It reduces an $n \times n$ symmetric matrix to a similar tridiagonal one by performing $n-2$ orthogonal transformations. This method finds a wide range of applications in linear algebra. The details of how the method works can be found in refs. [1,2]. Here we present the algorithm in MatLab language for given general $n \times n$ symmetric matrix A:

```
[n,n]=size(A);
for k=1:n-2
    X=A(:,k);
    ss=0;
    for j=k+1:n
    ss=ss+X(j)^2;
    end
```

```
        S=sign(X(k+1))*sqrt(ss);
        R=sqrt(2*(S+X(k+1))*S);
        W=X;
        W(1:k)=0;
        W(k+1)=W(k+1)+S;
        W=(1/R)*W;
        V=A*W;
        c=W'*V;
        Q=V-c*W;
        A=A-2*W*Q'-2*Q*W';
    end
```

**B. Eigenvalue degeneracy in square arrays**

The eigenstate degeneracy of uniform square arrays can be in general investigated by using the point group $D_4$ [3]. Other symmetry groups have to be considered for arrays having different topologies. However here, and since we focus on square arrays only, we discuss this feature by using an alternative straightforward algebraic method. At lasing threshold $g = \gamma$, equation (1) written in the rotating frame, takes the form: $i \frac{da_{m,n}}{dt} = \left( \kappa_{m,n}^{m-1,n} a_{m-1,n} + \kappa_{m,n}^{m+1,n} a_{m+1,n} \right) + \left( \kappa_{m,n}^{m,n-1} a_{m,n-1} + \kappa_{m,n}^{m,n+1} a_{m,n+1} \right)$. By using separation of variable $a_{m,n}(t) = \varphi_m(t)\eta_n(t)$ and substituting back, it is straightforward to show that:

$$\frac{1}{\varphi_m}\left\{ i\frac{d\varphi_m}{dt} - \left( \kappa_{m,n}^{m-1,n} \varphi_{m-1} + \kappa_{m,n}^{m+1,n} \varphi_{m+1} \right) \right\} + \frac{1}{\eta_n}\left\{ i\frac{d\eta_n}{dt} - \left( \kappa_{m,n}^{m,n-1} \eta_{n-1} + \kappa_{m,n}^{m,n+1} \eta_{n+1} \right) \right\} = 0 \quad \text{(B.1)}$$

In the most general case, the two terms in the brackets cannot be satisfied independently. However, for square arrays with identical rows and identical columns, i.e. $\kappa_{m,n}^{m\pm1,n} = \kappa_{m,n'}^{m\pm1,n'} \equiv T_m^{m\pm1}$ $\kappa_{m,n}^{m,n\pm1} = \kappa_{m',n}^{m',n\pm1} \equiv J_n^{n\pm1}$, the two terms become independent and we obtain:

$$i\frac{d\varphi_m}{dt} = T_m^{m-1}\varphi_{m-1} + T_m^{m+1}\varphi_{m+1} \qquad (B.2.a)$$

$$i\frac{d\eta_n}{dt} = J_n^{n-1}\eta_{n-1} + J_n^{n+1}\eta_{n+1} \qquad (B.2.b)$$

Evidently if $\mu_m$ and $\mu_n$ are the eigenvalues associated with equations (B.2.a) and (B.2.b), respectively (note that the indices of $\mu$ characterize the different eigenvalues and do not indicate array elements), i.e. $\varphi_m \propto \varphi_o \exp(-i\mu_m t)$ and $\eta_n \propto \eta_o \exp(-i\mu_n t)$, it follows that the eigenvectors of full system satisfy $a_{m,n} = \varphi_m \eta_n \propto \varphi_o \eta_o \exp[-i(\mu_m + \mu_n)t]$ and the associated eigenvalues are given by $\mu_{m,n} = \mu_m + \mu_n$. Degeneracies thus occur whenever $\mu_m + \mu_n = \mu_{m'} + \mu_{n'}$ for any integer indices $m, n, m'$ and $n'$.

**Square uniform arrays**

From the above analysis, it is clear that $\mu_{m,n} = \mu_{n,m}$. Also in the absence of any accidental degeneracy, the eigenvalues $\mu_{m,m}$ are unique. In addition, if one writes equations (B.2.a) and (B.2.b) in Hamiltonian forms, i.e. $i\frac{d\vec{\varphi}}{dt} = H_\varphi \vec{\varphi}$ and $i\frac{d\vec{\eta}}{dt} = H_\eta \vec{\eta}$, it is easy to show that $\{\sigma_z, H_{\varphi,\eta}\} = 0$ where $[\sigma_z]_{ij} = \delta_{ij}(-1)^{i+1}$ and the brackets $\{\ \}$ denote anti-commutation. In other words, the Hamiltonians $H_{\varphi,\eta}$ respect chiral particle-hole symmetry: each positive eigenvalue must be accompanied by a negative eigenvalue [4]. Consequently, the eigenspectrum is symmetric about zero. If the integer $N$ is even, the eigenvalues of $H_{\varphi,\eta}$ do not include any zero value and the null eigenvalues of the 2D system are only of the form $\mu_{m,-m}$. As a result, the system exhibits

$N$ unique eigenvalues of the form $\mu_{m,m}$, $N$ zero eigenvalues and doubly degenerate eigenvalues of the form $\mu_{m,n} = \mu_{n,m}$. The total number of distinct eigenvalues is thus given by $N + 1 + \dfrac{N^2 - 2N}{2} = \dfrac{N^2 + 2}{2}$. We illustrate this result by the chart shown in figure B.1 for the $4 \times 4$ array where degenerate eigenvalues having the same value are highlighted by the same color. Similar considerations apply to the case of odd value of $N$, except that we must take into account that here the particle-hole symmetry forces one of the eigenvalues of $H_{\varphi,\eta}$ to be zero and thus results in an additional accidental degeneracy. In this case, the total number of distinct eigenvalues turns out to be $\dfrac{N^2 + 1}{2}$. By applying these formulas for the two square arrays discussed in section two, we indeed find that the $3 \times 3$ and $4 \times 4$ arrays exhibit five and nine distinct eigenvalues, respectively.

We note that in the above analysis, we have assumed that apart from the zero eigenvalue dictated by the chiral symmetry in the case when $N$ is odd, no other accidental degeneracy arises. Figure. B.1 illustrates these degeneracy for the case of $4 \times 4$ array where only nine distinct eigenfrequencies exist.

|        | $\mu_2$           | $\mu_1$           | $-\mu_1$          | $-\mu_2$          |
|--------|-------------------|-------------------|-------------------|-------------------|
| $\mu_2$ | $2\mu_2$          | $\mu_1+\mu_2$     | $\mu_2-\mu_1$     | $0$               |
| $\mu_1$ | $\mu_1+\mu_2$     | $2\mu_1$          | $0$               | $-(\mu_2-\mu_1)$  |
| $-\mu_1$ | $\mu_2-\mu_1$    | $0$               | $-2\mu_1$         | $-(\mu_1+\mu_2)$  |
| $-\mu_2$ | $0$              | $-(\mu_2-\mu_1)$  | $-(\mu_1+\mu_2)$  | $-2\mu_2$         |

Figure. B.1 Eigenvalue structure of a $4 \times 4$ uniform square array. Nine distinct eigenfrequencies exist as highlighted by the different colors.

**Square bosonic arrays**

The above discussion applies equally to the square bosonic arrays introduced in section 3 in the main text. However, here in addition to the geometric induced degeneracies, accidental degeneracies also occur. In the other words, in bosonic arrays, the condition $\mu_m + \mu_n = \mu_{m'} + \mu_{n'}$ can be satisfied for a set of modes (characterized by the indices $m, n, m'$ and $n'$) that do not necessarily transform into one another under geometric operations such as reflection and rotation. In particular, due to the equidistant eigenvalue ladder of one dimensional bosonic arrays, the degeneracy condition $\mu_m + \mu_n = \mu_{m'} + \mu_{n'}$ in two dimensional configurations holds when $m + n = m' + n'$. By taking these accidental symmetries into account, we find that the total number of non-degenerate eigenstates in an $N \times N$ bosonic array is given by $2N - 1$. Figure. B.2 illustrates these degeneracies for the case of $4 \times 4$ bosonic array where only seven distinct eigenfrequencies exist.

|      | 3Δ  | Δ   | −Δ  | −3Δ |
| ---- | --- | --- | --- | --- |
| 3Δ   | 6Δ  | 4Δ  | 2Δ  | 0   |
| Δ    | 4Δ  | 2Δ  | 0   | −2Δ |
| −Δ   | 2Δ  | 0   | −2Δ | −4Δ |
| −3Δ  | 0   | −2Δ | −4Δ | −6Δ |

Figure. B.2 Eigenvalues of a $4 \times 4$ bosonic array where only seven distinct eigenfrequencies exist. Here we have an equidistant eigenvalue spectrum for $J_x$ laser array and $2\Delta$ corresponds to the eigenvalue ladder steps associated with the 1D Hamiltonians $H_{\varphi,\eta}$.

**C. Absence of degeneracy in 1D discrete systems**

As we have discussed in the main text, the eigenspectra of one dimensional systems do not exhibit any degeneracies. For completeness, we sketch here a simple proof of this known fact. A 1D discrete system is described by a tridiagonal matrix. For any given eigenvalue, the elements of the corresponding eigenvector can be expressed in terms of that eigenvalue, the matrix elements and the first component of that eigenvector which can be chosen arbitrarily. Now assume that there exist two different eigenvectors that correspond to the same eigenvalue. Due to the linearity of the problem, we can scale the first component of the second eigenvector to match that of the first. But as a result, the rest of the components of the eigenvectors will be equal after the scaling. This in turn means that the originally different eigenvectors were related by a constant multiplication factor and hence are basically the same. We note that these restrictions that lead to the impossibility of degeneracy in 1D systems are lifted in higher dimensions where degeneracy is allowed.